\documentclass[12pt,a4paper]{article}

\usepackage[dvips]{graphicx}

\usepackage{amsmath}
\usepackage{amsfonts}
\usepackage{bm}

\input{colordvi.tex}

\newcommand{\CC}{\mathbb{C}}
\newcommand{\ZZ}{\mathbb{Z}}

\newcommand{\tr}{{\rm tr}}

\newcommand{\ol}{\overline}

\newcommand{\wt}{\widetilde}
\newcommand{\wh}{\widehat}

\def\diag{\mathop{\rm diag}}

\newcommand{\rap}[2]
{\setbox1=\hbox{#1}%
\setbox2=\hbox to\wd1{\hss #2\hss}%
\mbox{\rlap{\box1}\box2}}

\begin{document}\begin{titlepage}
\title{
\begin{flushright}
\normalsize{TIT/HEP-619\\
August 2012}
\end{flushright}
       \vspace{2cm}
${\bm S}^3/\ZZ_n$ partition function and dualities
       \vspace{2cm}}
\author{
Yosuke Imamura\thanks{E-mail: \tt imamura@phys.titech.ac.jp}\quad
and\ \ Daisuke Yokoyama\thanks{E-mail: \tt d.yokoyama@th.phys.titech.ac.jp}
\\[30pt]
{\it Department of Physics, Tokyo Institute of Technology,}\\
{\it Tokyo 152-8551, Japan}
}
\date{}

\maketitle
\thispagestyle{empty}

\vspace{0cm}

\begin{abstract}
\normalsize
We investigate $\bm S^3/\ZZ_n$ partition function of $\mathcal N = 2$
supersymmetric gauge theories.
A gauge theory on the orbifold has degenerate vacua
specified by the holonomy.
The partition function is obtained by summing up
the contributions of saddle points with different holonomies.
An appropriate choice of the phase of each contribution
is essential
to obtain the partition function.
We determine the relative phases in the holonomy sum
in a few examples by using duality to
non-gauge theories.
In the case of odd $n$ the phase
factors can be absorbed by modifying a single function
appearing in the partition function.
\end{abstract}

\end{titlepage}

\tableofcontents
\section{Introduction}

For last few years,
there was large progress in supersymmetric field theories
in three-dimension.
Many physical quantities
in strongly coupled field theories have been computed by using localization method.
For example, partition functions of ${\cal N}=2$ supersymmetric field theories
in ${\bm S}^3$ \cite{Kapustin:2009kz,Jafferis:2010un,Hama:2010av} and
${\bm S}^2\times{\bm S}^1$ \cite{Kim:2009wb,Imamura:2011su} are computed exactly,
and they have been used to study non-perturbative aspects of
three-dimensional field theories,
such as dualities among three-dimensional field theories \cite{Kapustin:2010xq,Kapustin:2010mh,Kapustin:2011gh,Willett:2011gp}
and relation to M-theory via AdS/CFT correspondence \cite{Kim:2009wb,Drukker:2010nc,Herzog:2010hf}

In this paper, we investigate the partition function of three-dimensional
${\cal N}=2$ supersymmetric field theories in the orbifold ${\bm S}^3/{\ZZ}_n$ \cite{Gang:2009wy,Benini:2011nc}.
Due to the non-trivial homotopy of the orbifold,
$\pi_1({\bm S}^3/{\ZZ}_n)={\ZZ}_n$,
a gauge theory defined in it
has degenerate vacua specified by the holonomy associated
with the gauge symmetry.
Their contributions are summed up
to obtain the total partition function.
In general, the partition function of a Euclidean theory
is complex.
We usually focus only on its absolute value
and the phase is disregarded.
This is, however, not allowed when
we compute the partition functions of different sectors which are
summed up.
Even when we are interested only in the absolute value of the
total partition function, we need to care about the relative
phase of each contribution.
The purpose of this paper is to determine appropriate phase factors
in the holonomy sum in some gauge theories and look for a general rule for these phases.
We consider two gauge theories which are known to have dual field theories without
vector multiplet.
On one side of the dualities, in the non-gauge theories,
we can compute the absolute value of the partition function
up to overall constant factor independent of parameters.
By comparing the partition functions of gauge and non-gauge theories in each dual pair,
we infer the relative phases in the holonomy sum in the gauge theories.

This paper is organized as follows.
In \S\ref{pf.sec},
we summarize a general formula of the orbifold partition function.
It is written by using orbifold extension of the double sign function,
which we denote by $s_{b,h}(z)$.
In \S\ref{dual.sec}, we consider two dual pairs
and determine the phase factors in the holonomy sum in the gauge theories so that the sum agrees with
the partition function of the dual non-gauge theories.
We find that when the order $n$ of the orbifold group is odd
the phase factor is absorbed in the definition of the
function $s_{b,h}(z)$.
In \S\ref{der.sec} we consider more dual pairs which are derived
from one of the dualities studied in \S\ref{dual.sec}.
The last section is devoted to the conclusions and discussions.

\section{The ${\bm S}^3/{\ZZ}_n$ partition function}\label{pf.sec}
\subsection{The ${\bm S}^3$ partition function}

Let us first summarize the partition function on ${\bm S}^3$ without
orbifolding \cite{Kapustin:2009kz,Jafferis:2010un,Hama:2010av,Hama:2011ea,Imamura:2011wg}.
We consider the squashed ${\bm S}^3$ with the metric
\begin{equation}
ds^2=r^2\left[(\mu^1)^2+(\mu^2)^2+\frac{1}{v^2}(\mu^3)^2\right],
\end{equation}
where $\mu^a$ ($a=1,2,3$) are the left-invariant differentials on
${\bm S}^3\sim SU(2)$.
We set $r=1$ in the following.
$v$ is the squashing parameter.
We also define parameters $u$ and $b$ by
\begin{equation}
u=\pm\sqrt{v^2-1},\quad
b=\frac{1+iu}{v},
\end{equation}
for later use.
The isometry of the round sphere $SU(2)_L\times SU(2)_R$ is broken by the squashing
to
\begin{equation}
SU(2)_L\times U(1)_r.
\label{su2u1}
\end{equation}
We consider an ${\cal N}=2$ supersymmetric gauge theory on this manifold.
The supercharges belong to the representation ${\bm 2}_0$ of (\ref{su2u1}).

A general formula for the partition function for the squashed sphere
is given in \cite{Imamura:2011wg}.
The same partition function is first obtained in
\cite{Hama:2011ea} for a different deformation of ${\bm S}^3$.
See also
\cite{Dolan:2011rp,Gadde:2011ia,Imamura:2011uw}
for its relation to 4d superconformal index.
The formula is
\begin{equation}
Z=\int[d\lambda]e^{-S_0(\lambda)}Z^{\rm 1-loop},
\label{s3formula}
\end{equation}
where $Z^{\rm 1-loop}$ is the one-loop determinant
\begin{equation}
Z^{\rm 1-loop}=
\frac{\prod_{\alpha\in\Delta} s_{b}\left(\alpha(\lambda)-\frac{i}{v}\right)}
{\prod_I s_{b}\left(\rho_I(\lambda)-\frac{i(1-\Delta_I)}{v}\right)}.
\label{zs3a}
\end{equation}
The index $I$ labels chiral multiplets,
and $\rho_I$ and $\Delta_I$ are the weight vectors and the Weyl weights of
the chiral multiplets.
The integration variable $\lambda$ is an element of the Cartan subalgebra
of the gauge group $G$ parameterizing the Coulomb branch.
If $G$ is $U(N)$ or product of $U(N)$,
the integration measure $[d\lambda]$
is defined by
\begin{equation}
[d\lambda]=\frac{1}{|W|}\prod_{a=1}^{{\rm rank} G}d\lambda_a,
\label{s3measure}
\end{equation}
where $|W|$ is the order of the Weyl group of $G$,
and $\lambda_a$ are the diagonal components in the
fundamental representation of $\lambda$.
Different normalizations are also used in the literature.
The normalization
(\ref{s3measure})
is chosen so that
theories in the dual pairs we will consider in \S\ref{dual.sec}
have the same partition functions.\footnote{
The normalization of $\lambda$ we use in this paper
is different from that in \cite{Imamura:2011wg}.
$\lambda$ here is related to the constant mode $\sigma_0$ of
scalar field defined in \cite{Imamura:2011wg} by $\lambda=r\sigma_0/v$.}

$S_0(\lambda)$ is the classical action.
If the theory has the Chern-Simons term
\begin{align}
S_{\rm CS}&=\frac{ik}{4\pi}\int\tr_{\rm fund}\left(AdA-\frac{2i}{3}A^3\right),
\end{align}
and the Fayet-Iliopoulos term
\begin{align}
S_{\rm FI}&=-\frac{\zeta}{2\pi}\int\sqrt{g}D_{U(1)} d^3x,
\end{align}
then the supersymmetric completion of these actions
contributes to the classical action by
\begin{equation}
S_0(\lambda)=\pi i k\tr_{\rm fund}(\lambda^2)
+2\pi i\zeta\lambda_{U(1)}.
\end{equation}

$s_b(z)$ is the double sine function
defined by\footnote{
$s_b(z)$ is directly related to the non-compact quantum dilogarithm $\varphi_b(z)$ \cite{Faddeev:1994fw,Faddeev:1995nb,Faddeev:2000if} by
$s_b (z) = \exp[-\pi i(\frac{z^2}{2}+\frac{b^2+b^{-2}}{24})] \varphi_b(z)$.
See also \cite{Kharchev:2001rs, Bytsko:2006ut} for more details on $s_b(z)$.
}

\begin{equation}
s_b(z)=
\prod_{p,q=0}^\infty
\frac{b\left(q+\frac{1}{2}\right)+b^{-1}\left(p+\frac{1}{2}\right)-iz}
{b\left(p+\frac{1}{2}\right)+b^{-1}\left(q+\frac{1}{2}\right)+iz}.
\label{doublesine}
\end{equation}
The formula
(\ref{zs3a})
is derived by using localization,
which reduces
the path integral to Gaussian integral
for infinite number of non-zero modes and
finite dimensional
integral for zero modes.
The Gaussian integral is performed by using spherical harmonics expansion.
For a vector multiplet with weight vector $\alpha$,
the Gaussian integral of modes with specific $SU(2)_R$ quantum numbers $(j,m)$ gives
\begin{equation}
\frac{2j-2imu-iv\alpha(\lambda)}{2j+2+2imu+iv\alpha(\lambda)}.
\label{jmvector}
\end{equation}
Notice that $m$ dependence appears only when ${\bm S}^3$ is squashed.
For round sphere with $u=0$, there is no $m$ dependence
due to the unbroken $SU(2)_R$ symmetry.
If we set
\begin{equation}
j=\frac{p+q}{2},\quad
m=\frac{p-q}{2},
\label{mnpq}
\end{equation}
(\ref{jmvector}) becomes the factor
in the definition (\ref{doublesine}) of the double sine function $s_b(z)$
with argument
$z=\alpha(\lambda)-i/v$,
and by taking the product over $p$ and $q$,
we obtain the double sine function appearing in the numerator in
(\ref{zs3a}).
The denominator in (\ref{zs3a}) also arises from the Gaussian integral
of non-zero modes of chiral multiplets.

Note that variables $p$ and $q$ appear differently
in the numerator and the denominator in (\ref{doublesine}).
In the numerator $p$ and $q$ appear in the coefficients of $b^{-1}$ and $b$,
respectively,
and in the denominator, the relation is reversed.
Because they are dummy variables
we can exchange them
in the numerator or in the denominator
so that they appear in the same manner.
However,
the relation to the $SU(2)_R$ quantum numbers (\ref{mnpq})
holds only when we write the infinite product as in
(\ref{doublesine}).
This becomes important when we consider the orbifolding by
${\ZZ}_n\subset SU(2)_R$ in the next subsection.

From the definition (\ref{doublesine}) we can easily show the
following relations for $s_b(z)$.

\begin{itemize}
 \item

Self-duality and reflection property
\begin{equation}
s_b(z)
=s_{b^{-1}}(z)
=\frac{1}{s_b(-z)}.
\label{self}
\end{equation}

\item
Functional equations
\begin{align}
\frac
{s_b(z+\frac{ib}{2})}
{s_b(z-\frac{ib}{2})}
&=\frac{1}{2\cosh(\pi b z)},\nonumber\\
\frac
{s_b(z+\frac{ib^{-1}}{2})}
{s_b(z-\frac{ib^{-1}}{2})}
&=\frac{1}{2\cosh(\pi b^{-1} z)},\nonumber\\
\frac
{s_b(z+\frac{i}{v})}
{s_b(z-\frac{i}{v})}
&=\frac{1}{[2\sinh(\pi b z)][2\sinh(\pi b^{-1} z)]}.
\label{funceq}
\end{align}

\end{itemize}

\subsection{${\ZZ}_n$ orbifolding}
We consider
the left-invariant orbifold ${\bm S}^3/{\ZZ}_n$ with
${\ZZ}_n\subset U(1)_r\subset SU(2)_R$.
The partition function on the orbifold
is obtained in \cite{Gang:2009wy}
for theories without matter fields in
a general Lens space $L(p,q)$ without squashing.
It is extended to theories with chiral multiplets
in background with nontrivial squashing parameter in
\cite{Benini:2011nc}.
Our orbifold corresponds to $L(n,-1)$.

Because supercharges are $U(1)_r$ neutral,
the orbifolding by ${\ZZ}_n\subset U(1)_r$
does not break any supersymmetry,
and we can define ${\cal N}=2$ supersymmetric theories on the orbifold.
A gauge theory in this orbifold has degenerate vacua specified by the holonomy
\begin{equation}
m=\frac{n}{2\pi}\oint_C A,
\label{mdef}
\end{equation}
where $C$ is the generator of the fundamental group $\pi_1({\bm S}^3/{\ZZ}_n)={\ZZ}_n$.
The consistency to $nC=0$
requires $e^{2 \pi im}=1$.
(Note that we define $m$ with the factor $n$ in (\ref{mdef}).)
The holonomy can be turned on for both global and gauge symmetries.
The holonomy for gauge symmetries should be summed up in the path integral.
The partition function is given by
\begin{equation}
Z(m_{\rm global})=\sum_{m_{\rm local}}\int[d\lambda]e^{-S_0(\lambda,m)}
Z^{\rm 1-loop}(\lambda,m),
\label{znpartitionfunc}
\end{equation}
where $S_0(\lambda,m)$ and $Z^{\rm 1-loop}(\lambda,m)$ are
the classical action and the one-loop determinant.
The summation
is taken over
the holonomy associated with gauge symmetry,
which is denoted by
$m_{\rm local}$ in (\ref{znpartitionfunc}).
The holonomy for global symmetry $m_{\rm global}$ is not summed, and the
partition function $Z$ depends on $m_{\rm global}$.

The integration measure $[d\lambda]$ is defined by
\begin{equation}
[d\lambda]=\frac{1}{|W|}\prod_{a=1}^{{\rm rank} G}\frac{d\lambda_a}{n}.
\label{orbmeasure}
\end{equation}
We introduce the factor $1/n$ for each integration
variable for later convenience.

One may think that the classical action for ${\bm S}^3/{\ZZ}_n$ is
obtained by dividing that for ${\bm S}^3$ by $n$.
This naive expectation is not correct.
The classical action $S_0(\lambda,m)$ consists of two parts;
\begin{equation}
S_0^{{\bm S}^3/{\ZZ}_n}(\lambda,m)
=\frac{1}{n}S_0^{{\bm S}^3}(\lambda)-i\Phi(m).
\end{equation}
One is $1/n$ of the classical action for ${\bm S}^3$,
and has the same origin as the ${\bm S}^3$ case.
The other part comes from the Chern-Simons term.
Due to the non-trivial topology of ${\bm S}^3/{\ZZ}_n$,
the Chern-Simons term gives
non-vanishing contribution
even for a flat gauge connection \cite{Gang:2009wy,0209403,Griguolo:2006kp};
\begin{equation}
\Phi=\frac{\pi k}{n}\tr_{\rm fund}(m^2).
\label{csphase}
\end{equation}
This phase
plays an important role in dualities in ${\bm S}^3/{\ZZ}_n$.
The factor $e^{i\Phi}$ may be ill-defined
depending on the coefficient.
If $nk$ is odd,
the holonomies $m=\diag(\cdots,h,\cdots)$ and $m=\diag(\cdots,h+n,\cdots)$, which are identified in $\ZZ_n$,
give different phases.
We will meet such an ambiguity in the example in \S\ref{jy.sec},
and there we will give an additional rule to fix the ambiguity.

The one-loop partition function for the orbifold
can be obtained by projecting out
the factors
in (\ref{doublesine}) which originate from $\ZZ_n$-variant
modes.
Let $\varphi$ be
a field with a weight vector $\rho$.
On ${\bm S}^3$ it is Fourier expanded as
\begin{equation}
\varphi(\psi)=\sum_{m\in{\bm Z}/2}\varphi_m e^{im\psi},
\end{equation}
where $0\leq\psi<4\pi$ is the coordinate along the Hopf fiber
of ${\bm S}^3$, and $m$ is the $SU(2)_R$ magnetic quantum number.
After $\ZZ_n$ orbifolding, the field must
satisfy the
boundary condition
\begin{equation}
\varphi\left(\psi+\frac{4\pi}{n}\right)=e^{2\pi i\frac{\rho(h)}{n}}\varphi(\psi),
\end{equation}
and
only modes $\varphi_m$ with the index $m$ satisfying
\begin{equation}
2m=p-q=\rho\cdot h\mod n
\label{pqcond}
\end{equation}
survive after the orbifold projection.
We define $s_{b,h}(z)$ as the function obtained
from (\ref{doublesine})
by restricting the product over $(p,q)$ by
(\ref{pqcond}).
This restricted product is realized by substituting
\begin{equation}
p=np'+[k+h]_n,\quad
q=nq'+k,
\label{pqbypqk}
\end{equation}
to (\ref{doublesine}), and perform the product
with respect to non-negative integers $p'$ and $q'$, and $k=0,1,\ldots,n-1$.
$[m]_n$ represents the remainder when $m$ is divided by $n$.
It is convenient to introduce notation $\langle\cdots\rangle_n$ defined by
\begin{equation}
\langle m\rangle_n=\frac{1}{n}\left([m]_n+\frac{1}{2}\right)-\frac{1}{2}.
\end{equation}
This satisfies the relations
\begin{equation}
\langle m+an\rangle_n=\langle m\rangle_n \quad (a \in \ZZ),\quad
\langle -1-m\rangle_n=-\langle m\rangle_n.
\label{lrangrel}
\end{equation}
We rewrite the numerator in (\ref{doublesine}) as
\begin{align}
&b\left(p+\frac{1}{2}\right)+b^{-1}\left(q+\frac{1}{2}\right)-iz
\nonumber\\
&=n\left[b\left(p'
+\langle k+h\rangle_n
+\frac{1}{2}\right)+b^{-1}\left(q'+\langle k\rangle_n+\frac{1}{2}\right)-i\frac{z}{n}\right].
\end{align}
The denominator in (\ref{doublesine}) is
also rewritten in a similar way, and
we obtain
\begin{align}
s_{b,h}(z)
&=
\prod_{k=0}^{n-1}
\prod_{p',q'=0}^\infty
\frac{b(q'+\frac{1}{2})+b^{-1}(p'+\frac{1}{2})
+b\langle k\rangle_n
+b^{-1}\langle k+h\rangle_n
-i\frac{z}{n}}
{b(p'+\frac{1}{2})+b^{-1}(q'+\frac{1}{2})
+b\langle k+h\rangle_n
+b^{-1}\langle k\rangle_n
+i\frac{z}{n}}
\nonumber\\
&=
\prod_{k=0}^{n-1}
\prod_{p',q'=0}^\infty
\frac{b(q'+\frac{1}{2})+b^{-1}(p'+\frac{1}{2})
+b\langle k\rangle_n
+b^{-1}\langle k+h\rangle_n
-i\frac{z}{n}}
{b(p'+\frac{1}{2})+b^{-1}(q'+\frac{1}{2})
-b\langle k\rangle_n
-b^{-1}\langle k+h\rangle_n
+i\frac{z}{n}}.
\end{align}
In the second line we replaced $k$ in the denominator by $-1-k-h$
and used the second relation in
(\ref{lrangrel}).
In the final expression the product with respect to $p'$ and $q'$
has the same form as that in the definition
(\ref{doublesine})
of the double sine function
$s_b(z)$, and we obtain
\begin{equation}
s_{b,h}(z)=
\prod_{k=0}^{n-1}
s_b\left(
\frac{z}{n}
+ib\langle k\rangle_n
+ib^{-1}\langle k+h\rangle_n
\right).
\label{doublesine2}
\end{equation}
By definition,
the product of $s_{b,h}(z)$ over all $h$ reproduces the original double sine function;
\begin{equation}
s_b(z)=\prod_{h=0}^{n-1}s_{b,h}(z).
\end{equation}
The function $s_{b,h}(z)$ satisfies
the following formulae, which are analogs of (\ref{self}) and (\ref{funceq}).

\begin{itemize}
\item 

Self-duality and reflection property
\begin{equation}
s_{b,h}(z)
=s_{b^{-1},-h}(z)
=\frac{1}{s_{b,-h}(-z)}
=\frac{1}{s_{b^{-1},h}(-z)}.
\end{equation}

\item
Functional equations
\begin{align}
\frac
{s_{b,h+1}(z+\frac{ib}{2})}
{s_{b,h}(z-\frac{ib}{2})}
&=\frac{1}{2\cosh\left(\frac{\pi b z}{n}+\pi i\langle h\rangle\right)}
,\nonumber\\
\frac
{s_{b,h-1}(z+\frac{ib^{-1}}{2})}
{s_{b,h}(z-\frac{ib^{-1}}{2})}
&=\frac{1}{2\cosh\left(\frac{\pi b^{-1} z}{n}+\pi i\langle -h\rangle\right)}
,\nonumber\\
\frac{s_{b,h}(z+\frac{i}{v})}{s_{b,h}(z-\frac{i}{v})}
&=\frac{1}{
[2\sinh\left(\frac{\pi b z+\pi ih}{n}\right)]
[2\sinh\left(\frac{\pi b^{-1} z-\pi ih}{n}\right)]}.
\end{align}

\end{itemize}
The one-loop determinant for ${\bm S}^3/{\bm Z}_n$ is
obtained simply by replacing $s_b(z)$ in (\ref{zs3a}) by $s_{b,h}(z)$.
\begin{equation}
Z^{\rm 1-loop}(\lambda,m)
=\frac{\prod_{\alpha\in\Delta} s_{b,\alpha(m)}\left(\alpha(\lambda)-\frac{i}{v}\right)}
{\prod_I s_{b,\rho_I(m)}\left(\rho_I(\lambda)-\frac{i(1-\Delta_I)}{v}\right)}.
\label{lensz}
\end{equation}
The $1$-loop determinant
of vector multiplets can be rewritten in terms of elementary functions.
\begin{align}
&Z^{\rm 1-loop}_{\rm vector}(\lambda,m)
=\prod_{\alpha\in\Delta}s_{b,\alpha(m)}\left(\alpha(\lambda)-\frac{i}{v}\right)
\nonumber\\
&=\prod_{\alpha\in\Delta_+}
\left[2\sinh\frac{\pi}{n}\left(b\alpha(\lambda)+i\alpha(m)\right)\right]
\left[2\sinh\frac{\pi}{n}\left(b^{-1}\alpha(\lambda)-i\alpha(m)\right)\right].
\end{align}
When $b=1$, this agree with the partition function in
the lens space $L(n,-1)$ given in \cite{Gang:2009wy}.

\section{Dualities in ${\bm S}^3/{\ZZ}_n$}\label{dual.sec}
A gauge theory in ${\bm S}^3/{\ZZ}_n$
has degenerate vacua 
labeled by holonomies associated with the gauge symmetry.
The contributions of these vacua should
be summed up to obtain the total partition function.
In this section we consider two dual pairs
and confirm that
the partition functions of theories dual to each other
agree if appropriate phase factors are inserted in the holonomy sum.

\subsection{${\cal N}=2$ SQED and XYZ model}\label{n2mirror.sec}
We first consider the mirror symmetry between
an ${\cal N}=2$ SQED and the XYZ model \cite{deBoer:1997ka}.
On one side of the duality,
we consider ${\cal N}=2$ SQED with
two chiral multiplets $q$ and $\wt q$ with $U(1)$ charge $+1$ and $-1$, respectively.
We assume that $q$ and $\wt q$ have the Weyl weight $\Delta$.
The mirror theory, the XYZ model,
consists of three chiral multiplets
$Q$, $\wt Q$ and $S$
interacting through the superpotential
\begin{equation}
W=\wt QSQ.
\label{qsq}
\end{equation}
Although three fields $S$, $Q$, and $\wt Q$ in this model are
symmetric,
we treat $Q$ and $\wt Q$ as a quark and an antiquark,
and $S$ as a neutral field, because the mirror pair
considered here is a special case of a series of mirror pairs,
which are studied in \ref{nf.sec},
and
in a general mirror pair $Q$ and $\wt Q$ are replaced by
charged multiplets and $S$ by neutral ones.
By the operator relation $S=\wt q q$ and the marginality of the superpotential (\ref{qsq}),
we can determine the Weyl weights
of the fields in this theory as
\begin{equation}
\Delta_S=2\Delta,\quad
\Delta_Q=\Delta_{\wt Q}=1-\Delta.
\end{equation}
Although the correct value of $\Delta$ at the infra-red fixed point is
$\Delta=1/3$, the equality of partition functions holds regardless of
$\Delta$ \cite{Willett:2011gp}, and we leave $\Delta$ unfixed.

The global symmetry which is the same for these two theories
is $U(1)_V\times U(1)_A$.
The charge assignments are shown in Table \ref{table:1}.
\begin{table}[htb]
\caption{The charge assignment of the global symmetry $U(1)_V\times U(1)_A$
of SQED and XYZ model which are mirror to each other.
$m$ and $\wt m$ are the monopole and anti-monopole operators.}
\label{table:1}
\begin{center}
\begin{tabular}{cccccccccc}
\hline
\hline
         && $q$ & $\wt q$ & $m$ & $\wt m$ && $Q$ & $\wt Q$ & $S$ \\
\hline
$U(1)_V$ && $0$ & $0$ & $1$ & $-1$ && $1$ & $-1$ & $0$ \\
$U(1)_A$ && $1$ & $1$ & $0$ & $0$ && $-1$ & $-1$ & $2$ \\
\hline
\end{tabular}
\end{center}
\end{table}
We introduce real mass parameters $\zeta$ and $\mu$ for $U(1)_V$ and $U(1)_A$,
respectively.
$U(1)_A$ symmetry in SQED is the topological $U(1)$ symmetry
acting on monopole operators, and the corresponding mass parameter $\zeta$ is the Fayet-Iliopoulos parameter.
The ${\bm S}^3$ partition functions of these theories
are
\begin{align}
Z^{\rm SQED}&=
\int\frac{e^{-2\pi i \zeta\lambda}}{
s_b(\lambda+\mu-\frac{i(1-\Delta)}{v})
s_b(-\lambda+\mu-\frac{i(1-\Delta)}{v})
}d\lambda,\nonumber\\
Z^{\rm XYZ}&=
\frac{1
}
{
s_b(\zeta-\mu-\frac{i\Delta}{v})
s_b(-\zeta-\mu-\frac{i\Delta}{v})
s_b(2\mu-\frac{i(1-2\Delta)}{v})
}.
\end{align}

Because two theories are mirror to each other,
the partition functions should agree.
This agreement is confirmed by using the pentagon relation
of the double sine function \cite{Bazhanov:2007mh}\footnote{The pentagon relation usually
refers to the operator equation
$\varphi_b(\hat P)\varphi_b(\hat X)=\varphi_b(\hat X)\varphi_b(\hat X+ \hat P)\varphi_b(\hat P)$,
where $\hat X$ and $\hat P$ are operators satisfying $[\hat P, \hat X]=1/2 \pi i$.
This relation is proved in \cite{woronowicz,Faddeev:2000if}.
This operator equation is equivalent to (\ref{pentagon}) \cite{Faddeev:2000if,Ponsot:2000mt, Kashaevncqd}, which we refer to as the pentagon relation.}:
\begin{equation}
\int\frac{s_b(x+r)}{s_b(x+s)}e^{-2\pi i tx}dx
=e^{\pi it(r+s)}
\frac{s_b(t-\frac{r}{2}+\frac{s}{2}+\frac{i}{v})}
{s_b(t+\frac{r}{2}-\frac{s}{2}-\frac{i}{v})}s_b(r-s-\frac{i}{v}).
\label{pentagon}
\end{equation}
By substituting
\begin{equation}
x=\lambda,\quad
r=-\mu+\frac{i(1-\Delta)}{v},\quad
s=\mu-\frac{i(1-\Delta)}{v},\quad
t=\zeta,
\label{params}
\end{equation}
to the pentagon relation (\ref{pentagon}),
we obtain $Z^{\rm XYZ}=Z^{\rm SQED}$.
Note that for the agreement of the two partition functions,
the integration measure should be chosen as in (\ref{s3measure}).

Let us generalize this to the theories on the orbifold ${\bm S}^3/{\ZZ}_n$.
On the SQED side, we need to sum up the contribution
of $n$ saddle points specified by the holonomy $h$
of the $U(1)$ gauge symmetry.
We can also introduce holonomies
$h_V$ and $h_A$ for $U(1)_V$ and $U(1)_A$ global symmetries
as non-dynamical background gauge potentials.
Because $U(1)_V$ current in SQED is the field strength
of the dynamical gauge field $A$,
the $U(1)_V$ holonomy is realized by the Chern-Simons term
\begin{equation}
S=\frac{i}{2\pi}\int VdA,
\label{intvda}
\end{equation}
where $V$ is the non-dynamical $U(1)_V$ background gauge field.
In the orbifold ${\bm S}^3/{\ZZ}_n$, this term
gives rise to the non-trivial phase factor
\begin{equation}
\Phi=2\pi\frac{h_Vh}{n}.
\end{equation}
Taking account of this phase factor,
the partition function for each holonomy
is
\begin{eqnarray}
&&Z^{\rm SQED}(h,h_V,h_A)\nonumber\\
&&=\int_{-\infty}^\infty\frac{e^{-2\pi i\zeta\lambda/n}e^{2\pi i h_Vh/n}}
{s_{b,h_A+h}(\mu+\lambda-\frac{i(1-\Delta)}{v})
s_{b,h_A-h}(\mu-\lambda-\frac{i(1-\Delta)}{v})}
\frac{d\lambda}{n}.
\end{eqnarray}
On the other hand, the partition function of the XYZ model is
\begin{eqnarray}
&&Z^{\rm XYZ}(h_V,h_A)
\nonumber\\
&&=
\frac{1}{
s_{b,-h_A+h_V}(-\mu+\zeta-\frac{i\Delta}{v})
s_{b,-h_A-h_V}(-\mu-\zeta-\frac{i\Delta}{v})
s_{b,2h_A}(2\mu-\frac{i(1-2\Delta)}{v})}.
\label{xyz}
\end{eqnarray}
Naive expectation is that these are related by
\begin{equation}
Z^{\rm XYZ}(h_V,h_A)=\sum_{h=0}^{n-1}Z^{\rm SQED}(h,h_V,h_A).
\label{sigmaz0}
\end{equation}
This is actually the case when $h_A=0$.
We confirmed this relation numerically up to $n=10$.
Again, the choice of the integration measure
(\ref{orbmeasure}) is essential for the equality in (\ref{sigmaz0}).

The relation (\ref{sigmaz0}), however, does not hold
if we turn on the holonomy $h_A$ for
$U(1)_A$ symmetry.
Instead, we found that the relation
\begin{equation}
Z^{\rm XYZ}(h_V,h_A)=\sum_h \sigma(h,h_V,h_A) Z^{\rm SQED}(h,h_V,h_A)
\label{sigmaz}
\end{equation}
hold if we choose an appropriate sign function $\sigma(h,h_V,h_A)=\pm1$.
The analysis for $h_A=0$ implies
\begin{equation}
\sigma(h,h_V,0)=1.
\end{equation}
We can determine $\sigma(h,h_V,h_A)$ for general $h_A$ by the numerical analysis.
For $n=2,3,4$, we obtained
\begin{align}
&\sigma^{(2)}_1=\left(\begin{array}{rr}
-1 & 1 \\
1 & -1
\end{array}\right),\quad
\sigma^{(3)}_1
=\sigma^{(3)}_2
=\left(\begin{array}{rrr}
-1 &  1 &  1 \\
 1 & -1 & -1 \\
 1 & -1 & -1
\end{array}\right),\nonumber\\
&
\sigma^{(4)}_1
=\sigma^{(4)}_3
=\left(\begin{array}{rrrr}
-1 &  1 &  1 &  1\\
 1 & -1 & -1 & -1 \\
 1 & -1 & -1 & -1 \\
 1 & -1 & -1 & -1
\end{array}\right),\quad
\sigma^{(4)}_2=\left(\begin{array}{rrrr}
 1 & -1 &  1 & -1\\
-1 &  1 & -1 &  1 \\
 1 & -1 &  1 & -1 \\
-1 &  1 & -1 &  1
\end{array}\right),
\end{align}
where we express the function in the matrix form
\begin{equation}
(\sigma^{(n)}_{h_A})_{h,h_V}=\sigma(h,h_V,h_A).
\end{equation}
We determined the signs up to $n=10$,
and found the general form
\begin{equation}
\sigma(h,h_V,h_A)=(-1)^{f(h_A)+g(h_A,h)+g(h_A,h_V)},
\label{eq:sign}
\end{equation}
where
\begin{equation}
f(h)=\min(|h+n{\bf Z}|),\quad
g(h,h')=\min(f(h),f(h')).
\label{fandg}
\end{equation}

\subsection{$SU(2)$ gauge theory and a chiral multiplet}\label{jy.sec}
As the second example,
we consider the duality proposed by Jafferis and Yin in \cite{Jafferis:2011ns}.
The theory on one side is $SU(2)$ Chern-Simons theory with level $k=1$
coupling to one adjoint chiral multiplet $\Phi$.
It is dual to the theory consisting of a single chiral multiplet $X$.
These theories have global symmetry $U(1)_A$ rotating $\Phi$ and $X$
with charges $1$ and $2$, respectively.

Let us first compute the ${\bm S}^3$ partition function of the $SU(2)$ theory.
We parameterize the $SU(2)$ Cartan algebra by
\begin{equation}
\lambda=xT_3,\quad
T_3=\left(\begin{array}{cc}
\frac{1}{2} & 0 \\
0 & -\frac{1}{2}
\end{array}\right),
\end{equation}
and
we adopt the integration measure
\begin{align}
 [d\lambda] &= \frac{dx}{2\sqrt 2},
\end{align}
where the factor $1/2$ comes from the order of the Weyl group of $SU(2)$,
and $1/\sqrt2$ from the normalization of the $SU(2)$ generators
$\tr T_aT_b=(1/2)\delta_{ab}$.
The classical value of the
Chern-Simons term with level $k=1$
is
\begin{equation}
S_0=\pi i\tr(\lambda^2)=\frac{\pi i}{2}x^2.
\label{s0su2}
\end{equation}
The partition function of the $SU(2)$ theory is
\begin{equation}
Z^{SU(2)}=\int
\frac{e^{-\frac{\pi}{2}ix^2}s_b(x-\frac{i}{v})s_b(-x-\frac{i}{v})}
{s_b(x-\frac{i(1-\Delta)}{v})s_b(-\frac{i(1-\Delta)}{v})s_b(-x-\frac{i(1-\Delta)}{v})}
\frac{dx}{2\sqrt2},
\label{su2ons3}
\end{equation}
where we denote the Weyl weight of the adjoint chiral multiplet by $\Delta$.
If we turn on the real mass parameter $\mu$ for $U(1)_A$
the weight $\Delta$ is replaced by $\Delta -iv\mu$.

The dual theory contains a single chiral multiplet $X$.
This corresponds to the gauge invariant operator $\tr\Phi^2$
in the $SU(2)$ gauge theory,
and has Weyl weight $2\Delta$.
The ${\bm S}^3$ partition function is
\begin{equation}
Z^X=\frac{1}{s_b(-\frac{i(1-2\Delta)}{v})}.
\end{equation}
We can easily check numerically that
these two partition functions
coincide up to a phase factor.
\begin{equation}
Z^{SU(2)}=e^{i\phi}Z^X,\quad
\phi=-\pi\left(\frac{1}{4}+\frac{2\Delta+\Delta^2}{2v^2}\right).
\label{zsu2andzx}
\end{equation}
This relation is confirmed numerically in \cite{Jafferis:2011ns} and analytically in \cite{Agarwal:2012wd} for the round sphere.
The coincidence of the absolute value is due to our choice of the integration
measure.
In \cite{Jafferis:2011ns} different measure is used and extra numerical factor arises.
We do not argue about this point, and focus only on the phases.
For the round sphere, 
the phase factor
\begin{equation}
e^{i\phi}
=\exp\left[\pi i\left(\frac{1}{4}-\frac{(1+\Delta)^2}{2}\right)\right]
=\int e^{\pi i t^2-\sqrt 2\pi i(1+\Delta)t} dt,
\end{equation}
is
interpreted in \cite{Jafferis:2011ns} as the contribution of
a decoupled topological sector.
For squashed ${\bm S}^3$, there seems no such a simple explanation
for this factor.

We would like to extend this relation to the orbifolds.
In the introduction of holonomy,
we should note that the gauge group is, precisely speaking, not $SU(2)$ but $SU(2)/\ZZ_2=SO(3)$.
The allowed holonomies are
\begin{equation}
\exp\left(2\pi i\frac{h}{n}T_3\right),\quad
h=0,\ldots,n-1.
\label{su2holo}
\end{equation}
(If the gauge group were $SU(2)$, $2\pi$ in the exponent in
(\ref{su2holo})
should be replaced by $4\pi$.)
For the flat connection specified by the holonomy $h$,
the classical Chern-Simons action gives the phase factor
\begin{equation}
e^{i\Phi}=e^{\frac{\pi i}{2n}h^2}.
\label{su2phase}
\end{equation}
This is not well defined as a map from $\ZZ_n$ to $\CC$.
This gives different phases for
$h$ and $h+n$, which are identical in $\ZZ_n$.
We will fix this ambiguity later by an additional rule.

The orbifold partition function of the $SU(2)$ theory is
obtained
from the ${\bm S}^3$ partition function (\ref{su2ons3})
by
\begin{itemize}
\item replacing each $s_b(z)$ by $s_{b,h}(z)$ with an appropriate holonomy,
\item replacing the measure $dx$ by $dx/n$,
\item replacing the classical action $S_0$
in (\ref{s0su2})
by $S_0/n$,
\item and introducing the phase factor $e^{\pi ih^2/2n}$.
\end{itemize}
We obtain
\begin{equation}
Z^{SU(2)}(h,h_A)=\int
\frac{e^{\pi i\frac{h^2}{2n}}e^{-\frac{\pi}{2n}ix^2}s_{b,h}(x-\frac{i}{v})s_{b,-h}(-x-\frac{i}{v})}
{s_{b,h_A+h}(x-\frac{i(1-\Delta)}{v})s_{b,h_A}(-\frac{i(1-\Delta)}{v})s_{b,h_A-h}(-x-\frac{i(1-\Delta)}{v})}
\frac{dx}{2\sqrt2 n}.
\end{equation}
The partition function of the chiral multiplet $X$ is
\begin{equation}
Z^X(h_A)=\frac{1}{s_{b,2h_A}(-\frac{i(1-2\Delta)}{v})}.
\end{equation}

We consider two cases with even $n$ and odd $n$ separately.
Let us first consider the case with odd $n$.
In this case, (\ref{su2phase}) defines double-valued map from $\ZZ_n$ to $\CC$.
For $h$ and $h+n$, which are identified in $\ZZ_n$,
the phase factor takes different values whose
phases always differ by $\pi/2$.
We denote these two phase factors by $(e^{\frac{\pi i}{2n}h^2})_\pm$.
The subscript $\pm$ is chosen so that the two phases satisfy
$(e^{\frac{\pi i}{2n}h^2})_+=i(e^{\frac{\pi i}{2n}h^2})_-$.
Corresponding to these two choices of the phase factor, we define
two partition functions $Z_\pm^{SU(2)}(h,h_A)$.

We take the ansatz
\begin{equation}
\sum_{h=0}^{n-1}\sigma(h,h_A)e^{\mp\frac{\pi i}{4}}Z_\pm^{SU(2)}(h,h_A)=e^{i\phi}Z^X(h_A),
\label{zxzsu2}
\end{equation}
between the partition functions of the dual theories.
$\sigma(h,h_A)$ is an unknown phase function depending on the $SU(2)$ holonomy $h$ and
$U(1)_A$ holonomy $h_A$,
and $e^{i\phi}$ is a phase factor independent of holonomies.
The double signs on the left hand side are in the same order.
The factor $e^{\mp\frac{\pi i}{4}}$ is inserted to cancel the
difference of $Z^{SU(2)}_+$ and $Z^{SU(2)}_-$.
Although we can choose one of signs as a convention and
absorb this factor by $\sigma(h,h_A)$ or $e^{i\phi}$,
we separate this factor for later convenience.
We carried out the numerical analysis up to $n=29$,
and we found
\begin{align}
\sigma(h,h_A)&=(-1)^{g(h_A,h)} \exp\left[ i \pi \frac{f(h_A)(f(h_A)+n)}{2n} \right],\nonumber\\
\phi(h_A)&=-\pi
\frac{\Delta^2+2\Delta}{2nv^2},
\label{sigmaphidef}
\end{align}
make
the equation (\ref{zxzsu2}) hold,
where $f$ and $g$ are the functions defined in
(\ref{fandg}).

Let us turn to the case with even $n$.
In this case
we divide $n$ possible holonomies
to the $n/2$ satisfying
\begin{equation}
h-\frac{n}{2}\in 2\ZZ_n,
\label{evencond}
\end{equation}
and the others.
The phase factor
(\ref{su2phase}) is
well-defined for holonomies satisfying (\ref{evencond}),
while
(\ref{su2phase}) has
the sign ambiguity for the other holonomies.
With the numerical analysis up to $n=30$,
we found that $Z^X(h_A)$
can be given as a linear combination
of only $Z^{SU(2)}(h,h_A)$
with $h$ satisfying
(\ref{evencond}),
\begin{equation}
\sum_{h-n/2\in 2\ZZ_n}\sigma(h,h_A)\sqrt2Z^{SU(2)}(h,h_A)=e^{i\phi}Z^X(h_A),
\label{orbsu2xrel}
\end{equation}
where $\sigma(h,h_A)$ and $\phi(h_A)$ are functions defined in (\ref{sigmaphidef}).
Comparing this to (\ref{zxzsu2}),
we notice that the phase factor $e^{\pm\frac{\pi i}{4}}$
is replaced by $\sqrt2=e^{\frac{\pi i}{4}}+e^{-\frac{\pi i}{4}}$.
Although this factor
depends on the choice of the integration measure
and this may not have physical significance,
it may be interesting to discuss what this factor implies
under the assumption that our choice of the integration measure
is an appropriate one.
One possible interpretation is as follows.
In the theory of the chiral multiplet $X$, the $U(1)_A$ holonomy
appear only through $2h_A$.
When $n$ is even, there are two holonomies which gives the same $2h_A$.
Let $h_A$ be one of them, and $h_A'=h_A+n/2$ the other.
It is natural to sum up
the contribution of these two holonomies
on the $SU(2)$ theory side.
If we introduce different phase factors $e^{+\frac{\pi i}{4}}$ and $e^{-\frac{\pi i}{4}}$
for $h_A$ and $h'_A$ in this
summation, we obtain the following relation similar to (\ref{zxzsu2}).
\begin{align}
e^{i\phi}Z^X(h_A) = e^{i\phi}Z^X(h'_A) =
&\sum_{h-n/2\in 2\ZZ_n}\sigma(h,h_A)e^{\frac{\pi i}{4}}Z^{SU(2)}(h,h_A)
\nonumber\\
&+\sum_{h-n/2\in 2\ZZ_n}\sigma(h,h'_A)e^{-\frac{\pi i}{4}}Z^{SU(2)}(h,h_A').
\end{align}

\subsection{${\bm S}^3/\ZZ_{2k+1}$}
In the previous subsections,
we found that we need non-trivial
phase factors to match the partitions functions
of dual theories in two examples.
For odd $n$, in fact, we can express these phase factors
in a unified way.
Let us define $\sigma_h$ by
\begin{equation}
\sigma_h=(-1)^{[h]_n([h]_n-(-1)^{(n-1)/2})/2}.
\end{equation}
When $n$ is odd, this takes values $\pm1$ depending on $h\in\ZZ_n$.
We can represent $(-1)^{f(h)}$ and $(-1)^{g(h,h')}$
with this function by
\begin{equation}
(-)^{f(h)}=\sigma_{2h},\quad
(-)^{g(h,h')}=\sigma_{h+h'}\sigma_{h-h'}.
\end{equation}
Therefore, the sign function
(\ref{eq:sign}) in the first example can be given
as the product of five $\sigma_h$;
\begin{equation}
\sigma(h,h_V,h_A)
=\sigma_{h-h_A}\sigma_{h+h_A}
\sigma_{h_V+h_A}
\sigma_{h_V-h_A}
\sigma_{2h_A}.
\end{equation}
The indices of five $\sigma_h$ coincide up to sign
with the holonomy indices of the
functions $s_{b,h}(z)$ appearing in the mirror relation
(\ref{sigmaz}).
Because $\sigma_h=\sigma_{-h}$ and the sign of the index of $\sigma_h$ does not matter,
the phases can be absorbed into the
definition of the function $s_{b,h}(z)$.
Namely, if we define
$\wh Z^{\rm SQED}$ and $\wh Z^{\rm XYZ}$ from
$Z^{\rm SQED}$ and $Z^{\rm XYZ}$, respectively,
by replacing $s_{b,h}(z)$ in these partition functions
by $\wh s_{b,h}(z)$ defined by
\begin{equation}
\wh s_{b,h}(z)=\sigma_hs_{b,h}(z),
\label{shat}
\end{equation}
the relation
\begin{equation}
\wh Z^{\rm XYZ}(h_V,h_A)=\sum_{h=0}^{n-1}
\wh Z^{\rm SQED}(h,h_V,h_A)
\end{equation}
holds without the extra sign factors.
This is actually the case in the second example.
Because the phase function can be written as
\begin{equation}
\sigma(h,h_A)=\sigma_{h_A+h}\sigma_{h_A-h}\exp\left[ i \pi \frac{f(h_A)(f(h_A)+n)}{2n} \right],
\end{equation}
$\wh Z^{SU(2)}$ and $\wh Z^X$ defined with $\wh s_{b,h}(z)$
satisfy the relation
\begin{equation}
\sum_{h=0}^{n-1}e^{\mp\frac{\pi i}{4}}\wh Z_\pm^{SU(2)}(h,h_A)=\omega(h_A) e^{i\phi}\wh Z^X(h_A),
\end{equation}
where $\omega(h_A)$ is a certain factor depending only on $h_A$.

In the two examples,
we found that if we replace $s_{b,h}(z)$ by $\wh s_{b,h}(z)$ the duality relations
hold without introducing non-trivial relative phase factors
in the holonomy sums.
This is simple enough for us to expect that this rule is universal.
It would be interesting to check whether this rule really holds
for other examples of dual pairs.

\section{Derived dualities}\label{der.sec}
In this section we discuss three more dualities
which can be derived
from the mirror symmetry studied in \S\ref{n2mirror.sec}.

\subsection{${\cal N}=4$ SQED and hypermultiplet}\label{hyp.subsec}
It is known that the $\mathcal N =4$ SQED with one flavor
is mirror to a hypermultiplet \cite{Intriligator:1996ex}.
This mirror pair is obtained from the ${\cal N}=2$ mirror pair
in \S\ref{n2mirror.sec} by adding
a chiral multiplet $\wt S$ on the both sides of the duality.
On the SQED side, the new chiral multiplet $\wt S$ couples to the
system through the superpotential $W=\wt q\wt Sq$.
This corresponds to the mass term $W=\wt SS$ on the other side
of the duality, and we can integrate out $S$ and $\wt S$
to obtain the system with a hypermultiplet $(Q,\wt Q)$.
The global symmetry of the resulting mirror pair is $U(1)_V\times U(1)_A$
with the charge assignment summarized in Table \ref{table:2}.
We again introduce the mass parameters $\zeta$ and $\mu$ for $U(1)_V$ and $U(1)_A$, respectively.
\begin{table}[htb]
\caption{Global symmetries for $\mathcal N = 4$ SQED and the hypermultiplets. $m$ and $\wt m$ are again
(anti-)monopole operators.}
\label{table:2}
\begin{center}
\begin{tabular}{cccccccccc}
\hline
\hline
         && $q$ & $\wt q$ & $\wt S$ & $m$ & $\wt m$ && $Q$ & $\wt Q$ \\
\hline
$U(1)_V$ && $0$ & $0$ & $0$ & $1$ & $-1$ && $1$ & $-1$ \\
$U(1)_A$ && $1$ & $1$ & $-2$ & $0$ & $0$ && $-1$ & $-1$ \\
\hline
\end{tabular}
\end{center}
\end{table}
We denote the Weyl weights of $q$ and $\wt q$ by $\Delta$.
Then the Weyl weight of $\wt S$ is $1-2\Delta$.
The introduction of $\wt S$ changes the partition functions by
the factor
\begin{equation}
\frac{1}{s_{b,-2h_A}(-2\mu+\frac{i(1-2\Delta)}{v})}
=\textstyle{s_{b,2h_A}(2\mu-\frac{i(1-2\Delta)}{v})}.
\label{zwts}
\end{equation}
The partition function of two theories are given by
\begin{align}
&Z^{{\cal N}=4}(\zeta,\mu;h,h_V,h_A)\nonumber\\
&=
\int_{-\infty}^\infty\frac{e^{-2\pi i\zeta\lambda/n}e^{2\pi i h_Vh/n}}
{
\begin{array}{l}
s_{b,h_A+h}(\mu+\lambda-\frac{i(1-\Delta)}{v})\\
\times s_{b,h_A-h}(\mu-\lambda-\frac{i(1-\Delta)}{v})\\
\times s_{b,-2h_A}(-2\mu-\frac{i(2\Delta-1)}{v})
\end{array}
}
\frac{d\lambda}{n},
\\
&Z^{\rm hyper}(\zeta,\mu;h_V,h_A)
\nonumber\\
&=
\frac{1}{
s_{b,-h_A+h_V}(-\mu+\zeta-\frac{i\Delta}{v})
s_{b,-h_A-h_V}(-\mu-\zeta-\frac{i\Delta}{v})}.
\end{align}
Because the factor
(\ref{zwts}) does not depend on $h$,
it is rather trivial that the
partition functions match if we use the same sign function (\ref{eq:sign}) as
in the ${\cal N}=2$ case.
Namely, the following relation holds.
\begin{align}
 Z^{\rm hyper}(\zeta,\mu;h_V,h_A) &= \sum_{h=0}^{n-1}\sigma(h,h_V,h_A) Z^{{\cal N}=4}(\zeta,\mu;h,h_V,h_A).
 \label{hyperN4}
\end{align}

\subsection{${\cal N}=2$ SQED with $N_f\geq2$ and quiver gauge theory}\label{nf.sec}
The mirror symmetry between ${\cal N}=2$ SQED with $N_f\geq2$ and
a quiver gauge theory \cite{deBoer:1997ka}
can be derived from the ${\cal N}=4$ mirror symmetry in the previous subsection.
This fact is used
in \cite{Kapustin:2011jm}
to prove the partition functions of the mirror theories
coincide to each other in the case of round ${\bm S}^3$.
In this subsection, we generalize this to ${\bm S}^3/\ZZ_n$.

We start with $N_f$ copies of the mirror pairs constructed in
the previous subsection, and gauge
the diagonal subgroup $U(1)_{\rm diag}$ of
$N_f$ $U(1)_V$ symmetries.
Let $V_i$ ($i=1,\ldots,N_f-1$) be the background gauge fields corresponding to the
$U(1)_V$ symmetries.
The gauging of $U(1)_{\rm diag}$ is realized by the replacement
\begin{equation}
V_i\rightarrow V_i'+V_{\rm diag},
\label{vtovv}
\end{equation}
where $V_{\rm diag}$ is the dynamical gauge field of $U(1)_{\rm diag}$.
On one side of the duality, we have $N_f$ pairs of
the chiral multiplets $(Q_i,\wt Q_i)$ and
the $U(1)_{\rm gauge}$ vector multiplet.
These form ${\cal N}=2$ SQED with $N_f$ flavors.

\begin{figure}[ht]
 \begin{center}
  \includegraphics[width=9cm]{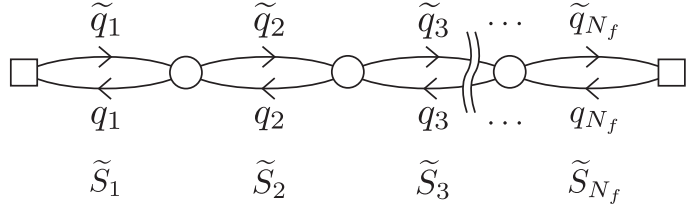}
  \caption{Quiver diagram of the mirror of $\mathcal N=2$ SQED with $N_f$ flavors.
  Circles and squares represent the $U(1)$ gauge groups and the $U(1)$ flavor symmetry, respectively.
  $q_i$ and $\wt q_i$ represented by arrows are bifundamental fields, and $\wt S_i$ are neutral fields.}
  \label{fig:001}
 \end{center}
\end{figure}

On the other side, we have $N_f$ copies of
SQED containing fields $(A_i,q_i,\wt q_i,\wt S_i)$ ($i=1,\ldots,N_f$)
and the $U(1)_{\rm gauge}$ vector multiplet $V_{\rm gauge}$.
These fields form a $U(1)^{N_f+1}$ gauge theory.
Because each copy of SQED has the Chern-Simons term (\ref{intvda}),
the gauging of $U(1)_{\rm gauge}$ by the replacement
(\ref{vtovv})
induces the Chern-Simons term
\begin{equation}
S=\frac{i}{2\pi}\int V_{\rm diag}dA_D,
\end{equation}
where $A_D=\sum_{i=1}^{N_f}A_i$ is the gauge field of the diagonal
subgroup $U(1)_D$ of the $N_f$ $U(1)$ gauge symmetries.
The equation of motion of $V_{\rm diag}$ gives the constraint
\begin{equation}
A_D=0
\label{adis0}
\end{equation}
on the gauge fields $A_i$.
We can solve this by
\begin{equation}
A_i=\wt A_i-\wt A_{i-1},
\label{anfm1}
\end{equation}
where $\wt A_i$ ($i=1,\ldots,N_f-1$) are independent dynamical gauge fields
and $\wt A_0=\wt A_{N_f}=0$.
As the result we have $U(1)^{N_f-1}$ quiver gauge theory
represented by the quiver diagram in Fig. \ref{fig:001}.

Let us confirm that the partition functions of the SQED and the $U(1)^{N_f-1}$ quiver gauge theory agree.
What is non-trivial is how we should choose the relative phases
in the holonomy sum for the newly introduced $U(1)_{\rm diag}$ gauge symmetry.
In the following, we will find that we need a non-trivial
sign factor depending on the $U(1)_{\rm gauge}$ holonomy
for the reduction of the gauge group from $U(1)^{N_f+1}$ to $U(1)^{N_f-1}$
works in the context of the ${\bm S}^3/\ZZ_n$ partition function.

Let us write down the orbifold partition functions.
Corresponding to (\ref{vtovv}) we replace
the $U(1)_V$ holonomies $h_V^i$ and $U(1)_V$ mass parameters $\zeta_i$
by
\begin{equation}
\zeta_i=\zeta_i'+x,\quad
h_V^i=h_V'^i+h',
\end{equation}
where $x$ and $h'$ are the $U(1)_{\rm diag}$ modulus
and the $U(1)_{\rm diag}$ holonomy.
On the side of $N_f$ pairs of chiral multiplets $(Q_i,\wt Q_i)$
we obtain
\begin{align}
&Z^{N_f}(\zeta_i',\mu_i;\vec h'_V,\vec h_A)\nonumber\\
&=
\sum_{h'=0}^{n-1}\tau(h',\vec h'_V,\vec h_A)
\int \frac{dx}{n}\prod_{i=1}^{N_f}Z^{\rm hyper}(x+\zeta_i',\mu_i;h'+h_V'^i,h_A^i)
\nonumber\\
&=
\sum_{h'=0}^{n-1}\tau(h',\vec h'_V,\vec h_A)
\int \frac{dx}{n}
\prod_{i=1}^{N_f}\frac{1}{s\cdots s},
\label{multiplavor}
\end{align}
where $s\cdots s$ represents the product of $s_{b,h}(z)$ coming from $Z^{\rm hyper}$.
We use arrows to represent sets of $N_f$ parameters;
$\vec h'_V=(h'^1_V,\ldots,h'^{N_f}_V)$ etc.
We introduced unknown sign function $\tau(h',\vec h'_V,\vec h_A)$
depending on the holonomies.

Application of the same prescription
on the side of $N_f$ copies of ${\cal N}=4$ SQED
gives the partition function
\begin{align}
&Z^{\rm quiver}(\zeta_i',\mu_i;\vec h'_V,\vec h_A)
\nonumber\\
&=
\sum_{h'=0}^{n-1}\tau(h',\vec h'_V,\vec h_A)
\int\frac{dx}{n}\prod_{i=1}^{N_f}
Z^{{\cal N}=4}(x+\zeta_i',\mu_i;h'+h_V'^i,h^i_A)
\nonumber\\
&=
\sum_{h'=0}^{n-1}\tau(h',\vec h'_V,\vec h_A)
\int\frac{dx}{n}
\nonumber\\&
\quad
\quad
\prod_{i=1}^{N_f}
\left[\sum_{h_i=0}^{n-1}
\sigma(h_i,h'+h_V'^i,h^i_A)\int_{-\infty}^\infty
\frac{d\lambda_i}{n}
\frac{
e^{-2\pi i(x+\zeta_i')\lambda_i/n}
e^{2\pi i (h'+h'^i_V)h_i/n}
}{s\cdots s}
\right].
\label{zquiver0}
\end{align}
Again, we represent the product of $s_{b,h}(z)$ coming from
$Z^{{\cal N}=4}$ by $s\cdots s$.
By definition this is the same as (\ref{multiplavor}).
What is non-trivial is if this partition function can be regarded as
that for quiver gauge theory with the reduced gauge group $U(1)^{N_f-1}$.
For this to be the case,
the summation over the holonomies $h'$ and $\vec h$ and
integral over the parameters $x$ and $\vec\lambda$ should
reduce to those over $N_f-1$ parameters corresponding to the
$N_f-1$ dynamical
gauge fields.

Let us first look at the $x$-integral in (\ref{zquiver0}).
The relevant part is
\begin{equation}
\int\frac{dx}{n}
\prod_{i=1}^{N_f}e^{-2\pi i(x+\zeta_i')\lambda_i/n}
=
\delta\left(\sum_{i=1}^{N_f}\lambda_i\right)
\exp\left(-\frac{2\pi i}{n}\sum_{i=1}^{N_f}\zeta_i'\lambda_i\right)
\label{deltadunc}
\end{equation}
The delta function
reduces the dimension of
the $\lambda_i$ integral by one,
and imposes the constraint
\begin{equation}
\sum_{i=1}^{N_f}\lambda_i=0.
\label{sumlambda}
\end{equation}
We can solve this by
\begin{equation}
\lambda_i=\wt\lambda_i-\wt\lambda_{i-1},\quad
\wt\lambda_0=\wt\lambda_{N_f}=0.
\end{equation}
The $N_f-1$ parameters $\wt\lambda_i$ correspond to the $N_f-1$ dynamical vector multiplets
introduced in
(\ref{anfm1}).

Concerning the summation with respect to $h'$,
the relevant part in (\ref{zquiver0}) is
\begin{align}
&
\sum_{h'=0}^{n-1}
\tau(h',\vec h'_V,\vec h_A)
\prod_{i=1}^{N_f}
\left(
\sigma(h_i,h'+h_V'^i,h^i_A)
e^{2\pi i (h'+h'^i_V)h_i/n}
\right)
\nonumber\\
&=
\sum_{h'=0}^{n-1}
\left(
\tau(h',\vec h'_V,\vec h_A)
\prod_{i=1}^{N_f}
\sigma(h_i,h'+h_V'^i,h^i_A)\right)
e^{\left(\frac{2\pi i}{n}h'\sum_{i=1}^{N_f}h_i\right)}
e^{\left(\frac{2\pi i}{n}\sum_{i=1}^{N_f}h'^i_Vh_i\right)}
\label{sumhp}
\end{align}
For (\ref{zquiver0}) to be regarded as the partition function of
the $U(1)^{N_f-1}$ quiver
gauge theory,
the condition
\begin{equation}
\sum_{i=1}^{N_f} h_i=0,
\label{condonhp}
\end{equation}
corresponding to (\ref{adis0})
and (\ref{sumlambda})
should arise.
Namely, the summation over $h'$
in (\ref{sumhp})
should give
the Kronecker's delta imposing the condition (\ref{condonhp}).
For this to happen
the sign factor in front of the exponentials in the summand must
be $h'$ independent.
Because $\tau(h',\vec h'_V,\vec h_A)$ is $h_i$-independent,
this is possible only when $h_i$ dependence and $h'$ dependence of
$\sigma(h_i,h'+h'^i_V,h_A^i)$ are factorized.
This is indeed the case as (\ref{eq:sign}) shows,
and we can make the coefficients in
(\ref{sumhp}) $h'$ independent
by choosing the function
$\tau(h',\vec h'_V,\vec h_A)$
as
\begin{equation}
\tau(h',\vec h'_V,\vec h_A)
=
\prod_{i=1}^{N_f}
(-1)^{f(h_A^i)+g(h_A^i,h'+h_V'^i)}.
\end{equation}
With this choice the summation over $h'$ gives
\begin{align}
&
\sum_{h'=0}^{n-1}
\tau(h',\vec h'_V,\vec h_A)
\prod_{i=1}^{N_f}
\left(
\sigma(h_i,h'+h_V'^i,h^i_A)
e^{2\pi i (h'+h'^i_V)h_i/n}\right)
\nonumber\\
&=
n\upsilon(\vec h,\vec h_A)
\delta\left(\sum_{i=1}^{N_f}h_i\right)
\exp\left(\frac{2\pi i}{n}\sum_{i=1}^{N_f}h'^i_Vh_i\right)
\label{kronecker}
\end{align}
where $\delta(*)$ in this equation is the Kronecker's delta,
and we defined the sign function
\begin{equation}
\upsilon(\vec h,\vec h_A)
=
\prod_{i=1}^{N_f}
(-1)^{g(h_A^i,h_i)}.
\end{equation}
The constraint imposed on $h_i$ by the Kronecker's delta
in (\ref{kronecker}) can be solved by
\begin{equation}
h_V'^i=\wt h_i-\wt h_{i-1},\quad
\wt h_i=\wt h_{N_f}=0.
\end{equation}
$\wt h_i$ are the holonomies of the $U(1)^{N_f-1}$ gauge symmetry.
We can rewrite
(\ref{zquiver0}) as
\begin{equation}
Z^{\rm quiver}(\vec h_V',\vec h_A)
=
\prod_{i=1}^{N_f-1}
\left(\sum_{\wt h_i}
\int\frac{d\wt\lambda_i}{n}\right)
\upsilon(\vec h,\vec h_A)\frac{
e^{-2\pi i \vec\zeta'\cdot\vec\lambda/n}
e^{2\pi i \vec h_V'\cdot\vec h/n}
}{s\cdots s}.
\end{equation}
This is precisely the partition function of the $U(1)^{N_f-1}$
quiver gauge theory.

\subsection{ABJM model and its mirror}
\begin{figure}[ht]
 \begin{center}
  \includegraphics[width=6cm]{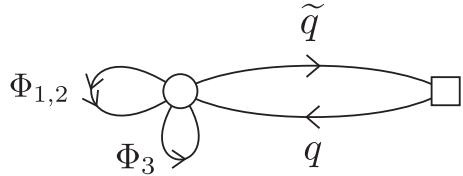}
  \caption{Quiver diagram of the mirror ABJM}
  \label{fig:002}
 \end{center}
\end{figure}

The ABJM model \cite{Aharony:2008ug} is an $U(N)\times U(N)$ Chern-Simons theory
with four bi-fundamental chiral multiplets $A_i$ and $B_i$ ($i=1,2$).
$A_i$ and $B_i$ belong to $({\bm N},\ol{\bm N})$ and $(\ol{\bm N},{\bm N})$, respectively.
The theory has the superpotential
\begin{equation}
W=\epsilon_{ik}\epsilon_{jl}\tr(A_iB_jA_kB_l),
\end{equation}
and all chiral multiplets have Weyl weight $1/2$.
When the Chern-Simons levels are $1$ and $-1$,
the ABJM model is known to be mirror to the ${\cal N}=4$ $U(N)$ gauge theory
with a fundamental hypermultiplet $(q,\wt q)$ and
an adjoint hypermultiplet $(\Phi_1,\Phi_2)$ \cite{Aharony:2008ug,Benini:2009qs,Kapustin:2010xq}.
In terms of ${\cal N}=2$ language this theory has the superpotential
\begin{align}
W&=\tr \Phi_3 \left( q \wt q +\left[ \Phi_1,\Phi_2 \right] \right),
\end{align}
where $\Phi_3$ is the chiral multiplet which form
the ${\cal N}=4$ vector multiplet together with the ${\cal N}=2$ vector multiplet.
The quiver diagram of this theory is shown in Fig. \ref{fig:002}.
The Weyl weights of the chiral multiplets are
\begin{equation}
\Delta_q=\Delta_{\wt q}=\Delta_{\Phi_1}=\Delta_{\Phi_2}=\frac{1}{2},\quad
\Delta_{\Phi_3}=1.
\end{equation}
These theories are expected to have $SO(8)$ global symmetry,
which are not manifest in the Lagrangians.
We only focus on the Cartan subalgebra
\begin{equation}
U(1)_A\times U(1)_B\times U(1)_T\times U(1)_R\subset SO(8),
\end{equation}
where $U(1)_R$ is the R-symmetry acting on the ${\cal N}=2$ supercharges.
The charge assignments in the two theories
are listed in Table \ref{table:001} and \ref{table:003}.
\begin{table}[htbp]
 \begin{center}
  \caption{Global symmetries of ABJM model} \vspace{10pt}
  \label{table:001}
  \begin{tabular}[tb]{c c rrrr}
   \hline   \hline
   && $A_1$ & $A_2$ & $B_1$ & $B_2$ \\
   \hline
   $U(1)_A$ && $+1$ & $-1$ & $0$ & $0$ \\
   $U(1)_B$ && $0$ & $0$ & $+1$ & $-1$ \\
   $U(1)_T$ && $+1$ & $+1$ & $-1$ & $-1$ \\ \hline
  \end{tabular}
 \end{center}
\end{table}
\begin{table}[htbp]
 \begin{center}
  \caption{Charge assignments in the mirror theory of the ABJM model.
  $m$ and $\wt m$ are monopole and anti-monopole fields.} \vspace{10pt}
  \label{table:003}
  \begin{tabular}[tb]{cc rrrrrrr}
   \hline \hline
   && $\Phi_1$ & $\Phi_2$ & $\Phi_3$ & $m$ & $\wt m$ & $q$ & $\wt q$ \\ \hline
   $U(1)_A$ && $+1$ & $-1$ & $0$ & $0$ & $0$ & $0$ & $0$ \\
   $U(1)_B$ && $0$ & $0$ & $0$ & $+1$ & $-1$ & $0$ & $0$ \\
   $U(1)_T$ && $+1$ & $+1$ & $-2$ & $0$ & $0$ & $+1$ & $+1$ \\ \hline
  \end{tabular} \\
\vspace{10pt}
 \end{center}
\end{table}
$U(1)_B$ in the mirror theory is a topological $U(1)$ symmetry coupling to monopole operators,
and the corresponding mass parameter is the Fayet-Iliopoulos parameter.

Although this mirror symmetry holds for an arbitrary rank $N$
of the gauge groups, we only consider Abelian ($N=1$) case for simplicity.
Let us first consider the orbifold partition function in the mirror theory.
When the gauge group is $U(1)$, the system consists of
two sectors decoupled from each other:
the neutral hypermultiplet and the ${\cal N}=4$ SQED with one flavor.
In the SQED we need to carry out
the holonomy sum to obtain the partition function,
and it is non-trivial how we should choose the relative phases.
We here take the sign function (\ref{eq:sign}) used in \S\ref{hyp.subsec}.
We will see shortly that the partition function obtained
with this phase factor is indeed reproduced in the ABJM model, too,
with an appropriate choice of the phases in the holonomy sum
in the ABJM model.
The partition function of the mirror theory is
\begin{equation}
Z^{\rm mirror}(h_A,h_B,h_T)
=\sum_h\sigma(h, h_B,h_T)Z_{\rm mirror}(h,h_A,h_B,h_T)
\label{zmirrortot}
\end{equation}
with the contribution of each holonomy sector
\begin{align}
&Z_{\rm mirror}(h,h_A,h_B,h_T)
\nonumber\\
 &= \frac{1}{
 s_{b,h_A+h_T} \left(m_A +m_T -\frac{i}{2v}\right)
 s_{b,-h_A+h_T} \left(-m_A +m_T -\frac{i}{2v}\right)
 s_{b,-2h_T} (-2 m_T)
 }
 \nonumber\\ & \hspace{20pt} \times
 \int\frac{dx}{n} \frac{e^{2 \pi i \frac{h h_B}{n}} e^{-2 \pi i m_B x} }
 {
 s_{b,h+h_T}(x +m_T -\frac{i}{2v})
 s_{b,-h+h_T}(-x +m_T -\frac{i}{2v})
 }\nonumber\\[10pt]
 &=
Z^{\rm hyper}(m_A,-m_T;h_A,-h_T)|_{\Delta=\frac{1}{2}}
 Z^{{\cal N}=4}(m_B,m_T;h,h_B,h_T)|_{\Delta=\frac{1}{2}},
\end{align}
where we introduce mass parameters $(m_A,m_B,m_T)$
and holonomies $(h_A,h_B,h_T)$ corresponding to the
flavor symmetry $U(1)_A\times U(1)_B\times U(1)_T$.
$h$ is the holonomy for the gauge symmetry.
With the relation (\ref{hyperN4}), we can rewrite
the partition function (\ref{zmirrortot}) as the
product of two $Z^{\rm hyper}$;
\begin{equation}
Z^{\rm mirror}(h_A,h_B,h_T)
=Z^{\rm hyper}(m_A,-m_T;h_A,-h_T)|_{\Delta=\frac{1}{2}}
Z^{\rm hyper}(m_B,m_T;h_B,h_T)|_{\Delta=\frac{1}{2}} .
\label{zmirror}
\end{equation}

On the ABJM side,
we need to sum up $n^2$ contributions
parameterized by a pair of holonomies $(h_1,h_2)$ for
the gauge group $U(1)_1 \times U(1)_2$.
The partition function of the sector specified by $(h_1,h_2)$ is
\begin{align}
&Z^{\rm ABJM}(h_1,h_2,h_A,h_B,h_T)
\nonumber\\
&=
e^{i\Phi(h_1,h_2)}\int
\frac{d\lambda}{n}
\frac{d\wt\lambda}{n}
\frac{
\exp\left[-\frac{\pi i}{n}(\lambda^2-\wt\lambda^2)\right]
}
{
\begin{array}{l}
s_{b,h_A+h_T+h_1-h_2}(m_A+m_T+\lambda-\wt\lambda-\frac{i}{2v})
\\\times s_{b,-h_A+h_T+h_1-h_2}(-m_A+m_T+\lambda-\wt\lambda-\frac{i}{2v})
\\\times s_{b,h_B-h_T-h_1+h_2}(m_B-m_T+\wt\lambda-\lambda-\frac{i}{2v})
\\\times s_{b,-h_B-h_T-h_1+h_2}(-m_B-m_T+\wt\lambda-\lambda-\frac{i}{2v})
\end{array}
}.
\label{abjmzholo}
\end{align}
A question is if it is possible to choose an appropriate
phases in the holonomy sum.
The answer is rather simple.
We do not need any non-trivial phases in this sum.
Let us confirm this by summing up
(\ref{abjmzholo}) over holonomies $h_1$ and $h_2$.
If we define $h_{12}\equiv h_1-h_2$
and replace $h_1$ by $h_{12}+h_2$,
$h_2$ appears only in the phase factor
\begin{equation}
\Phi
=\frac{\pi}{n}(h_1^2-h_2^2)
=\frac{\pi}{n}(h_{12}^2+2h_2h_{12}).
\end{equation}
The summation with respect to $h_2$ gives non-vanishing result
only when $h_{12}=0$,
and we obtain
\begin{align}
&Z^{\rm ABJM}(h_A,h_B,h_T)
=\sum_{h_1,h_2=0}^{n-1} Z^{\rm ABJM}(h_1,h_2,h_A,h_B,h_T)
\nonumber\\&
=\sum_{h=0}^{n-1} Z^{\rm ABJM}(h,h,h_A,h_B,h_T)
=n Z^{\rm ABJM}(0,0,h_A,h_B,h_T)
\nonumber\\
&=
n\int
\frac{d\lambda}{n}
\frac{d\wt\lambda}{n}
\frac{
\exp\left[-\frac{\pi i}{n}(\lambda^2-\wt\lambda^2)\right]
}
{
\begin{array}{l}
s_{b,h_A+h_T}(m_A+m_T+\lambda-\wt\lambda-\frac{i}{2v})
\\\times s_{b,-h_A+h_T}(-m_A+m_T+\lambda-\wt\lambda-\frac{i}{2v})
\\\times s_{b,h_B-h_T}(m_B-m_T+\wt\lambda-\lambda-\frac{i}{2v})
\\\times s_{b,-h_B-h_T}(-m_B-m_T+\wt\lambda-\lambda-\frac{i}{2v})
\end{array}
}.
\end{align}
We can easily perform the integral and have
\begin{align}
&Z^{\rm ABJM}(h_A,h_B,h_T)\nonumber\\
&=
\frac{
1
}
{
\begin{array}{l}
s_{b,h_A+h_T}(m_A+m_T-\frac{i}{2v})s_{b,-h_A+h_T}(-m_A+m_T-\frac{i}{2v})
\\\times
s_{b,h_B-h_T}(m_B-m_T-\frac{i}{2v})s_{b,-h_B-h_T}(-m_B-m_T-\frac{i}{2v})
\end{array}
}
\nonumber\\[10pt]
&=
Z^{\rm hyper}(m_A,-m_T;h_A,-h_T)|_{\Delta=\frac{1}{2}}
Z^{\rm hyper}(m_B,m_T;h_B,h_T)|_{\Delta=\frac{1}{2}} .
\label{zabjm}
\end{align}
This result precisely agrees with
the partition function of the mirror theory (\ref{zmirror}).

\section{Conclusions and discussions}
We investigated relative phases in the holonomy sum,
which is necessary to obtain the partition functions of gauge theories
in ${\bm S}^3/\ZZ_n$.
We used dualities between gauge theories and non-gauge theories to determine the phases.

We first consider mirror symmetry between ${\cal N}=2$
SQED with one flavor and the XYZ model
containing three chiral multiplets.
We showed that with the appropriate choice of the phases in the holonomy sum
the partition functions of these theories coincides.
Furthermore, we found that when $n$ is odd, the
phase factor is absorbed by the redefinition of
the single function $s_{b,h}(z)$, the orbifold extension of the double sine function.
We also consider the duality
between $SU(2)$ gauge theory and a chiral multiplet
proposed by Jafferis and Yin.
We could again find phase factors which makes the duality relation hold.
When $n$ is odd the phases are absorbed by
redefining the function $s_{b,h}(z)$
in the same way
as in the first example.

We also confirmed that it is possible to find phase factors in three more examples of dual pairs,
which are derived from the mirror symmetry we studied first.

When $n$ is odd, in all these examples, the phase factors can be absorbed by the definition of
the function $s_{b,h}(z)$.
This fact strongly suggests that the modified function $\wh s_{b,h}$ in (\ref{shat}) always
gives a ``correct'' partition function in some sense.
It would be interesting to check whether the modified function gives the same partition functions
for theories in dual pairs we did not studied in this paper.

For the purpose of obtaining non-trivial evidences for dualities,
it is desirable that we first fix the relative phases in the holonomy sum
in each theory without relying on dualities.
This should be possible at least for saddle points belonging to the same component
of the configuration space.
In this case we should obtain the relative phases by analyzing carefully the behavior
of the integration measure
in the path integral under adiabatic deformations connecting saddle points.
It is also interesting to search for guiding principles to fix the ambiguity
for the relative phases for the contribution of topologically disconnected sectors.


The large $N$ limit of the partition function of the ABJM model on the oribfold
is considered in \cite{Alday:2012au},
and the coincidence of the free energy calculated from
the ABJM model and that from the gravity dual
is confirmed.
Though many sectors with different holonomies contribute to the partition function
the authors found that in the large $N$ limit it is dominated by
the specific contribution with a certain holonomy configuration
and the relative phases are not important.
However, when one consider the next leading order of $1/N$
the other contributions become significant, therefore,
the phase plays an important role there.

\section*{Acknowledgments}
We are grateful to M.~Honda for valuable discussions.
The work of Y.I. was partially supported by Grant-in-Aid for Scientific Research
(C) (No.24540260), Ministry of Education, Science and Culture, Japan.
D.Y. acknowledges the financial support from the Global Center of Excellence Program by MEXT,
Japan through the “Nanoscience and Quantum Physics” Project of the Tokyo Institute of Technology.

\end{document}